\documentclass[10pt,a4paper,showpacs,jcp,twocolumn,superscriptaddress]{revtex4-1}
\usepackage[utf8]{inputenc}
\usepackage{amssymb}
\usepackage{mathrsfs}
\usepackage{stmaryrd}
\usepackage{graphicx}
\usepackage{color}
\usepackage{epstopdf} 
\usepackage{fancyhdr}
\newcommand{\mb}{\mathbf}
\newcommand{\mc}{\mathcal}

\newcommand{\tb}{\textbf}
\newcommand{\mr}{\mathrm}
\newcommand{\figref}[1]{Fig.\ \ref{#1}}

\newcommand{\RE}[1]{\textcolor{black}{#1}}
\newcommand{\ud}{\mathrm{d}}

\newcommand{\df}{\delta}

\begin{document}

\title{Anomalous Critical Slowdown at a First Order Phase Transition
in Single Polymer Chains}

\author{Shuangshuang Zhang}
\affiliation{Department of Physics, Beijing Normal University, Beijing 100875, China}
\affiliation{Graduate School Of Excellence Materials Science in Mainz, Staudingerweg 9, D-55128 Mainz, Germany}

\author{Shuanhu Qi}
\email{qish@uni-mainz.de}
\affiliation{Institut f\"{u}r Physik, Johannes Gutenberg-Universit\"{a}t Mainz, Staudingerweg 9, D-55099 Mainz, Germany}

\author{Leonid I. Klushin}
\affiliation{Department of Physics, American University of Beirut, P. O. Box 11-0236, Beirut 1107 2020, Lebanon, and Institute of Macromolecular Compounds,  Russian Acad. Sci. Bolshoy 31, 199004 St. Petersburg, Russia }

\author{Alexander M. Skvortsov}
\affiliation{Chemical-Pharmaceutical Academy, Professora Popova 14, 197022 St. Petersburg, Russia}

\author{Dadong Yan}
\affiliation{Department of Physics, Beijing Normal University, Beijing 100875, China}

\author{Friederike Schmid}
\affiliation{Institut f\"{u}r Physik, Johannes Gutenberg-Universit\"{a}t Mainz, Staudingerweg 9, D-55099 Mainz, Germany}

\begin{abstract}

Using Brownian Dynamics, we study the dynamical behavior of a polymer grafted
onto an adhesive surface close to the mechanically induced
adsorption-stretching transition.  Even though the transition is first order,
(in the infinite chain length limit, the stretching degree of the chain jumps
discontinuously), the characteristic relaxation time is found to grow according
to a power law as the transition point is approached. We present a dynamic
effective interface model which reproduces these observations and provides an
excellent quantitaive description of the simulations data. The generic nature
of the theoretical model suggests that the unconventional mixing of features
that are characteristic for first-order transitions (a jump in an order
parameter) and features that are characteristic of critical points (anomalous
slowdown) may be a common phenomenon in force-driven phase transitions of
macromolecules.

\end{abstract}

\maketitle

\subsection{Introduction}

Phase transitions have been recognized to be among the most fascinating
phenomena in physics since the days of van der Waals, Boltzmann, and Gibbs
\cite{Phase_book1,phase_book2}. In recent years, (pseudo)phase transitions in
molecules have received increasing attention in biophysics and materials
science, as they provide the physical basis for important biological processes
\cite{unzip_DNA1,unzip_DNA2,DNA,semiflexible,desorption_Vilgis,desorption_Vilgis2}
and can be exploited for nanomaterial design~\cite{sensors1, sensors2,
switch_lett, smart_Q,sommer1,sommer2}. The interest in single macromolecules is
also spurred by advances in experimental techniques~\cite{AFM, tweezers}, which
facilitate the manipulation of single molecules.

Phase-transition-like phenomena at the single molecule level often have unusual
features \cite{UPT_R1,tension}.  The so-called ``adsorption-stretching''
transition is a noticeable example, where an end-grafted chain initially
adsorbed onto an adhesive surface desorbs due to a tension force acting on the
free end. Analytical theory and Monte Carlo simulations have shown that this
transition is first order, but nevertheless displays features that are
typically associated with continuous phase transitions.  In the limit of
infinite chain length, the order parameter (the height of the free end) jumps
discontinuously, and the heat capacity has a delta-function-like singularity,
indicating that the transition is first order.  However, the distribution of
the order parameter is always unimodal and there are no metastable states
\cite{unusual1,unusual2}. Furthermore, analytical theory predicts that the
order parameter fluctuations should show an anomalous, power-like
pre-transitional growth \cite{UPT_first_order}.  These features are typical of
a second order transition.  

The presence of large fluctuations at the adsorption-stretching transition
point suggests that the relaxational dynamics for single macromolecules might
slow down accordingly. Dynamical slowdown is common close to second order phase
transitions, such as, e.g., 
driven desorption transition of free polymer chains on adhesive substrates
\cite{slowdown1,slowdown2}, but it would be rather untypical close to a first
order transition. It should have a severe impact on the kinetics of processes
that rely on mechanically driven desorption transitions.

The purpose of the present work is to present a systematic study of the
static and dynamic behavior in such a system. We use Brownian dynamics (BD)
simulations to investigate a single polymer chain grafted onto an adhesive
surface in the vicinity of its adsorption-stretching transition point. The
model does not account for hydrodynamic and entanglement effects, but it does
capture the essential physics of the transition. Our simulations confirm the
predicted increase of order parameter fluctuations at the transition and
show that the relaxation dynamics close to the transition point slows down
anomalously according to a power law.  To our best knowledge, this is the first
time that critical slowdown was observed at a first order phase transition in
dynamical simulations. Furthermore, we present a simple effective model that
captures both the static and dynamic properties of this unusual phase
transition and may help to understand other, similar, molecular phase
transitions.

\section{Model description and Brownian dynamics}

We consider a coarse-grained polymer chain of $N$ beads connected via Gaussian
springs. One end of the chain is grafted to an impenetrable substrate located
at $z=0$, which otherwise imposes an attractive potential on all polymer beads.
We use periodic boundary conditions along the $x$ and $y$ directions and
impenetrable boundaries at $z=0$ and $z=L_z$ in a simulation box of size
$L_\mathrm{x}=L_\mathrm{y}= 4 \sqrt{N}$ and $L_\mathrm{z}=N$.  Here and
throughout this paper, lengths are expressed in units of the statistical
segment length $a$, energies in units of $k_B T$, and times in units of
$\zeta_0 a^2$, where $\zeta_0$ is the friction coefficient for each bead. In
these units, the Hamiltonian of the system in a discretized form is given by 
%
\begin{equation} \label{eq:H} 
\mc H=\frac{3}{2}\sum_{j=2}^N\Big(\mb R_j-\mb R_{j-1}\Big)^2+\frac{v}{2}\sum_\alpha\hat\rho^2_\alpha-FZ_1+\sum_{j=1}^{N-1}U_\mr a(Z_j),
\end{equation}
in which $\alpha$ denotes the indexes of mesh points.  The four terms represent
the elastic energy, the pairwise excluded volume interactions, \RE{the
potential energy associated with the end force, and the attractive interaction
with the substrate, respectively.} Here $\mb R_\mathrm{j}$ denotes the location
of the $j$th bead, $Z_\mathrm{j}$ the corresponding z-component (with $Z_\mr
N=0$), and non-bonded interactions are formulated in terms of the bead density
operator $\hat\rho_\alpha=\hat{\rho} (\mb{r}_\alpha) =\frac{1}{\Delta V}\sum_j
g(|\mb R_j-\mb r_\alpha|)$ \cite{SCF1,CG_MC}, and $\Delta V$ is the volume of
one mesh cell, $g$ being an assignment function depending on the distance
between the bead and the mesh point. We choose the function $g$ such that the
fraction assigned to a given mesh point is proportional to the volume of a
rectangle whose diagonal is the line connecting the particle position and the
mesh point on the opposite side of the mesh cell. This is the so called
Cloud-in-Cells (CIC) scheme \cite{CIC,PtoM,Milano}. We set the cloud/cell size
$\Delta V=1$ and z-components of vortices located at $z=0.5 + \RE{m}$ ($\RE{m}
\in \mathbb{N}$).  To simplify the notation for the coordinate of the free end
which is one of the main quantities of interest, in the rest of the paper we
drop the subscript: $Z=Z_1$.  The excluded volume parameter $v$ is set to $v=1$
such that the grafted polymer is in good (implicit) solvent. \RE{The surface
interaction potential is defined through
$U_\mathrm{a}(\mathbf{r})=-\varepsilon\min(1,3/2-z)$ for $0\le z\le 3/2$, and
$U_\mathrm{a}(\mathbf{r})=0$ otherwise.  Here, $\varepsilon>0$ is the energy
gain if a monomer is in contact with the substrate. Thus the potential equals
$-\varepsilon$ at $0\le z\le 1/2$, and then linearly approaches zero at
$z=3/2$.} We note that $\mathcal{H}$ is a continuous function of the bead
positions $R_\mathrm{j}$, since the density operator $\hat{\rho}(\mb{r})$ is a
continuous function of the $\mb{R}_\mathrm{j}$. The bead positions evolve
according to the equations of overdamped Brownian dynamics,
$\dot{\mb{R}_\mathrm{j}}=-{\partial \mathcal {H}}/{\partial
\mb{R}_\mathrm{j}}+\sqrt{2}\mb{f}_\mathrm{r}$, where $\mb{f}_\mathrm{r}$ is an
uncorrelated and Gaussian distributed random force with mean zero and variance
$\langle f_\mathrm{r\alpha}(t)
f_\mathrm{r\beta}(t')\rangle=\delta_{\alpha\beta}\delta(t-t')$
($\alpha,\beta=x,y,z$). The time step in the simulation is chosen $\delta t
=0.005$. Quantities of interest are extracted from the trajectories of the free
end bead $j=1$. For details of the BD simulation scheme, we refer to the
Supplementary material.

\begin{figure}[t]
\centering
\includegraphics[scale=0.3]{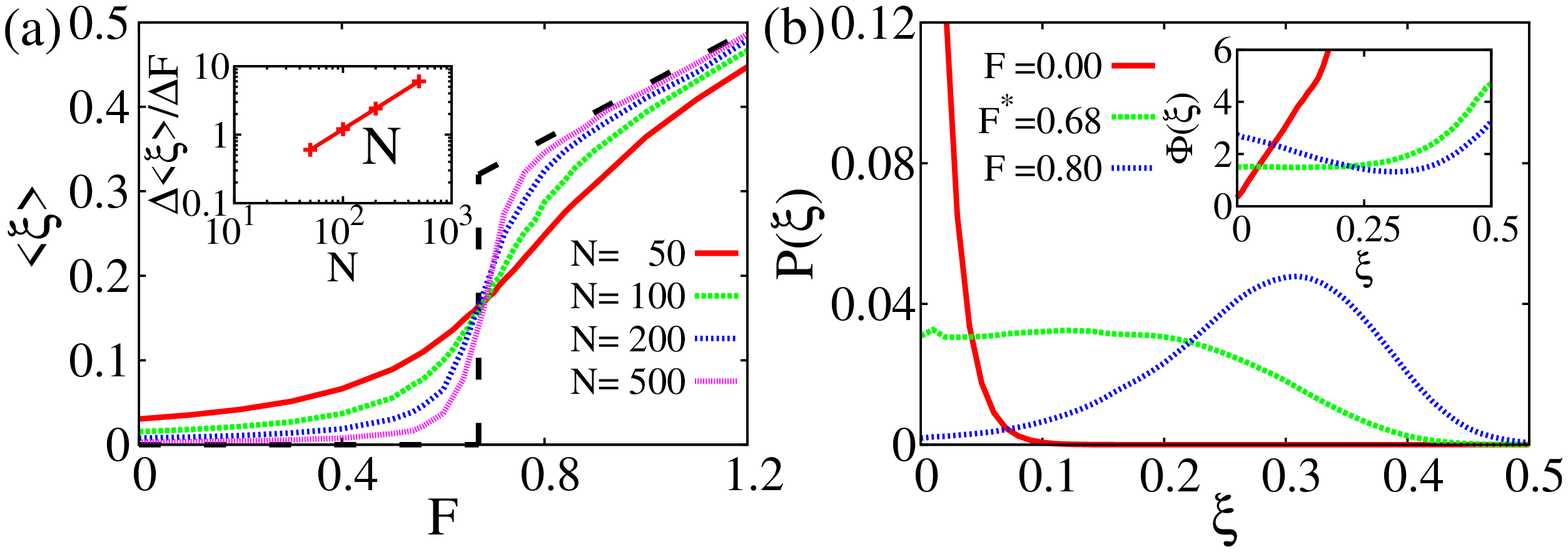}
\caption{ (a) Stretching degree $\langle \xi \rangle$ as a function of 
stretching force $F$ for chain lengths $N=50, 100, 200, 500$ 
at adsorption strength $-\varepsilon=0.80$. The dashed line represents the asymptotic limit $N\to\infty$.
Inset shows the maximum slope, which is proportional to $N$.
(b) Probability distribution of the order parameter $\xi$ at chain
length $N=100$ in the adsorbed state $F=0.00$, at the transition point 
$F^*=0.68$, and in the stretched state $F=0.80$. Inset shows the 
corresponding Landau free energy curves $\Phi(\xi)$.
}
\label{ZendvsFepsilon0.80}
\end{figure}

\begin{figure}[t]
\centering
\includegraphics[scale=0.35]{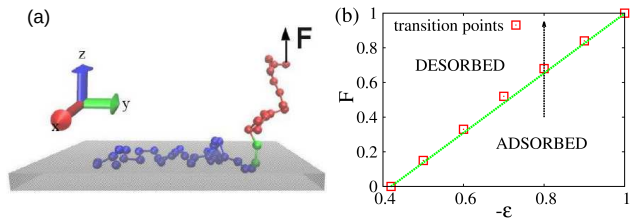}

\caption{ (a) Snapshot from BD simulation of the polymer chain at the
transition point ($-\varepsilon=0.80, F=0.68$), where the probability distribution
$P(\xi)$ exhibits a broad plateau. The gray region represents the attractive
surface layer. Blue beads are adsorbed, red ones belong to stretched
chain parts, and green ones are at the interface between these two states.
(b) Critical tensile force $F^{*}$ as a function of surface potential strength 
$\varepsilon$, which obeys a power law 
$F^{*}\propto (|\varepsilon|-|\varepsilon_\mathrm{c}|)^{x}$ with
$\varepsilon_\mathrm{c}=-0.42\pm0.02$, and $x= 1.0\pm 0.03$ (green line).}
\label{phasegraph}

\end{figure}

\section{Characterization of the phase transitions}

We begin with identifying the
order of the transition by examining the conformational properties of the
single chain, which can be characterized by the distribution of the free end.
Specifically, we choose the stretching degree $\xi= Z/N$ as the order
parameter,  where $ Z$ is the distance between the free end and the
substrate.  \figref{ZendvsFepsilon0.80}(a) shows the profiles of the mean order
parameter $\langle \xi \rangle$ as a function of the control parameter, the
pulling force $F$, at adsorption strength $-\varepsilon=0.8$ for several finite
chain lengths. Here $\langle\cdots\rangle$ denotes ensemble averages.  The
curves of different $N$ intersect at almost the same point, which hence
identifies the transition force $F^*$. In the vicinity of the transition, the
profile become sharper with increasing $N$. The maximum slope of each curve
increases linearly with increasing $N$ \figref{ZendvsFepsilon0.80}(a), inset).
Hence, in the limit of $N\to\infty$, a discontinuous abrupt jump of $\langle
\xi \rangle$ at the transition point can be expected (see the dashed line
in \figref{ZendvsFepsilon0.80}(a)).  This is a clear signal
of a first order transition, which is smoothened in simulations due to the 
finite chain lengths.

However, the order parameter distribution $P(\xi)$, shown in
\figref{ZendvsFepsilon0.80}(b), features a single broad peak at the transition
point $F^*=0.68$ and not a double peak structure as one would expect for a
first order transition. These observations are consistent with theoretical
predictions for ideal non-reversal random walks~\cite{unusual1} and for
self-avoiding chains~\cite{unusual2} as well as with simulation results for
spring-bead chains with hard core interactions \cite{unusual2}.  The
corresponding Landau free energy, which is defined as $\Phi(\xi)=-\ln P(\xi)$,
has only one minimum, there are no metastable states
(\figref{ZendvsFepsilon0.80} (b), inset). At the transition, the Landau free
energy has a wide plateau region stretching from $\xi=0$ to $\xi \sim 0.2$, and
there is no sign of an energy barrier between coexisting adsorbed and stretched
states. This can be explained from the fact that in this single-chain system,
the interface between stretched and absorbed chain sections comprises only a
few monomers (see \figref{phasegraph}(a)), hence the interfacial free
energies are negligibly small.

\begin{figure}[t]
\centering
\includegraphics[scale=0.45]{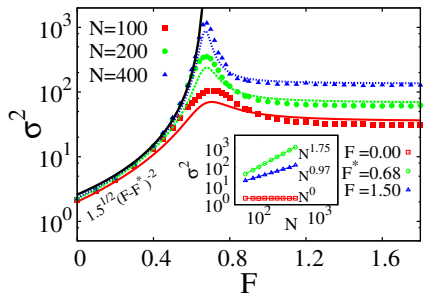}

\caption{Fluctuations $\sigma^2$ of the free end position $Z$ vs stretching
force $F$ for chain lengths $N=100,200,400$ at adsorption strength
$-\varepsilon=0.80$. Symbols are simulation data, lines correspond to the
analytical prediction, \RE{Eqs.\ (\ref{eq:sigma2}-\ref{eq:sigma22}). The thick
black line indicates the asymptotic power law divergence $(F-F^*)^{-2}$ upon
approaching $F^*$ from the adsorbed state} Inset shows scaling of $\sigma^2$
with $N$ in the adsorbed state ($F = 0$), at the transition ($F=F^*$), and in
the stretched state ($F=1.5$).  } \label{VZ}

\end{figure}

The phase diagram of the system is shown in \figref{phasegraph}(b) in the plane
of stretching force $F$ vs.\ adsorption strength $(- \varepsilon)$. A line of
order transitions $F^*(\varepsilon)$ separates the desorbed state from the
adsorbed state, which ends in a critical point at $(F^*,\varepsilon) =
(0,\varepsilon_c)$ with $\varepsilon_c = 0.42 \pm 0.02$. This point corresponds
to the thermodynamically driven adsorption transition, which is governed by the
competition between the attractive adsorption energy and \RE{the conformational
entropy loss for polymer chains near the impenetrable wall. The critical
adsorption point by definition refers to the thermodynamic limit $N\to\infty$,
and in this limit does not depend on whether the chain is grafted or not.} The
first order desorption line roughly follows $F^*\propto
(|\varepsilon|-|\varepsilon_\mathrm{c}|)^x$ with $x = 1.00 \pm 0.03$.

Next we examine the static behavior of the chain if the transition line is
crossed at fixed adsorption strength. We set $- \varepsilon=0.80$ (following
the black dashed arrow in \figref{phasegraph}(b)), and investigate the behavior
of the free end fluctuation normal to the substrate, $\sigma^2\equiv\langle
Z^2\rangle-\langle Z\rangle^2$.  \figref{VZ} displays $\sigma^2$ as a function
of the stretching force $F$. In the adsorbed state, when the stretching force
is small ($F<0.5$), $\sigma^2$ is largely independent of the chain length and
increases monotonically with increasing $F$.  In the stretched state at large
$F$, $F>1.2$, $\sigma^2$ is almost independent of $F$, but increases linearly
with the chain length.  It roughly obeys $\sigma^2 \approx N/3$, which is the
behavior of a deformed Gaussian coil. \RE{The BD model we utilize here does not
include the finite extensibility effect, as seen from the Hamiltonian of
Eq.(\ref{eq:H}). Thus the fluctuations become Gaussian fluctuation at strong
forces when the excluded volume effects become negligible.} As the transition
point is approached from \RE{below}, $\sigma^2$ first increases sharply and
then saturates at a finite value due to the finite chain length. The extremely
large values of $\sigma^2$ at the transition point simply reflect the fact that
the chain can assume adsorbed, stretched, and intermediate states with almost
equal probability. However, the behavior of $\sigma^2$ upon approaching the
transition point is unconventional. Usually, fluctuations exhibit a
$\delta$-peak at first order transitions, and power law divergences at second
order transitions.  Here the phase transition is first order, but the
fluctuations nevertheless diverge according to a power law, $\sigma^2 \sim
\frac{1}{(F-F^*)^2}$, independent of chain length. \RE{The divergence is
cut off at $F^*$ due to the finite length of the chains.}
This is analogeous to the behavior predicted analytically for ideal chains
\cite{UPT_first_order,note1}.  Right at the transition, $\sigma^2$ was
predicted to scale as $\sigma^2 \propto N^2$ with the chain length
\cite{unusual1, unusual2, UPT_first_order}, which is also in rough agreement
with our simulation results ($\sigma^2 \propto N^{1.75}$ according to
\figref{VZ}, inset).  \RE{The absolute value of $\sigma^2$ at the transition is
slightly underestimated by the theory. This is because the theory is based on
an ideal chain model, and in real chains, the fluctuations are further enhanced
due to excluded volume interactions.}

\section{Effective interface model}

Based on the observations above, we will now
develop a simple model which gives insight into the physical origin of the
anomalous behavior at this molecular phase transition, and which will provide a
starting point for discussing the dynamical properties at the transition.  Let
us first examine once more the snapshot of a chain at the transition point,
shown in \figref{phasegraph}(a). An adsorbed and a stretched block
coexist within the same molecule. The interfacial region connecting the two
blocks consists of only one or two segments (marked green), hence the excess
energy required for creating the interface is close to negligible.
Nevertheless, it is important to note that at most {\em one} such interface can
exist in the system. This is because the stretching force acting on the chain
end is transmitted along the stretched block up to the ``interface'', i.e., the
first contact with the surface, and then transferred to the substrate. It
does not directly influence the conformations of the adsorbed block. This
distinguishes the present system from conventional one dimensional systems such
as the one dimensional Ising model, where phase transitions are suppressed
because they are filled with many interfaces. Here, we have only one interface
separating the stretched and adsorbed block, which is free to move 
along the chain.

This motivates the construction of a dynamic effective interface model, which
describes the system in terms of the position $n$ of the interface within the
chain. Here $n$ is the label of the segment that establishes the first contact
with the substrate, counted from the free end: thus it has the meaning of the tail length. The interface is close to the
graft point ($n \to N$) in the stretched state and close to the end point ($n
\to 1$) in the adsorbed state, and at the transition, it moves abruptly from
one end to the other. The corresponding effective interface Hamiltonian has the
following generic form:
\begin{equation}
 \label{eq:Heff} 
 \mathcal{H}_\mathrm{eff}(n) = V_\mr l(n) + V_\mr r(N-n) -  \delta \: n,
\end{equation}where $\delta\propto (F-{F^*})$ is the difference between the adsorption and the
stretching free energies per monomer, while $V_\mr l(n)$ and $V_\mr r(n)$ are the left and right
repulsive potentials which ensure that $n$ stays within its bounds ($0 < n < N$). A hard-wall boundary is nearly exact for a strong pulling force whereby the stretched tail is not affected directly by the presence of an impenetrable substrate. However, it has a disadvantage that the minimum always resides at one of the boundaries which \RE{does} not allow for a simple analytical treatment of the dynamic behavior. We utilize instead a symmetric weakly diverging logarithmic potential, $V_\mr l(n)=V_\mr r(n)=-A\log(n)$, where $A$ is a numerical prefactor, leading to a very similar behavior at large $N$ with the benefit of generating a smooth minimum. In order to put further discussion on a quantitative basis we choose the value of $A$ to obtain the best match with the hard boundary model \RE{and} with the known results of the more sophisticated statistical theory \cite{unusual1,unusual2,UPT_first_order}. The minimum of $\mc H_\mr{eff}(n)$ defines the most probable tail length 
and is attained at
\begin{equation}
\label{eq:tail_length}
\bar n=N\left(\frac{1}{2}-\frac{A}{N\delta}\RE{\pm}\sqrt{\frac{1}{4}+\left(\frac{A}{N\delta}\right)^2}\right).
\end{equation}\RE{which has two branches, i.e., the positive and negative branch defined by the sign before the square root term. In the pre-transition region where $\delta<0$, we take the negative branch, while in the post-transition region where $\delta>0$, we take the positive branch. At $\delta=0$, we have $\bar n=N/2$. The most probable tail length obtained in the large $N|\delta|$ limit is $\bar n=N-A/\delta$ for $N\delta\gg 1$, and it is $\bar n=-A/\delta$ for $N\delta\ll -1$.}
It is clear that $\bar n(\delta)$ approaches a step function in the limit $N\to\infty$, $\bar n=N\Theta(\delta)$ 
leading to the observed jump in the order parameter characteristic of the first-order transition. The fluctuation of $n$ 
about $\bar n$ are evaluated as
\begin{equation}
\label{eq:sigma_n}
\sigma^2_\mr n=(V''_\mr l(\bar n)+V''_\mr r(N-\bar n))^{-1}=\frac{1}{A}\frac{\bar n^2(N-\bar n)^2}{\bar n^2+(N-\bar n)^2}.
\end{equation}The maximum fluctuations are obtained at the transition point, $\delta=0$, with $\bar n=N/2$ 
and $\sigma^2_\mr n=\frac{N^2}{8A}$, while far from the transition point, $N|\delta|\gg1$, the fluctuations decay 
as $\sigma^2_\mr n\simeq\frac{A}{\delta^2} $
On the other hand, the corresponding fluctuations for the hard boundary model are 
 $\sigma_\mr n^2=\frac{N^2}{12}$ at $\delta=0$ and $\sigma_\mr n^2\simeq\frac{1}{\delta^2}$ for $ N|\delta|\gg 1$, see Supplementary material.
It is clear that the choice of $A=3/2$ fixes the correct value of fluctuations at maximum, while $A=1$ properly represents 
 behavior away from the transition. To provide for the overall best match we choose the geometric mean value $A=\sqrt{3/2}$ hereafter.

To calculate the fluctuations $\sigma^2$ of the end position shown in
\figref{VZ}, we must combine these results with the statistical properties of
the desorbed tail. We assume that excluded volume effects are suppressed by
stretching and describe the stretched tail as a deformed Gaussian coil of $n$
segments, hence the end position $Z$ is Gaussian distributed with mean
$\bar{Z}_1(n) = n F/3$ and variance $\sigma_c^2(n) = n/3$. With these
assumptions, the full expression for the end fluctuations contains two
contributions,
\begin{equation}\label{eq:sigma2}
\sigma^2(Z)=\sigma_1^2(Z)+\sigma_2^2(Z)
\end{equation}
the first one describing the end fluctuations of the tail at the fixed (most
probable) tail length, $\sigma_1^2(Z)=\bar n/3$, and the second
$\sigma_2^2=\sigma^2_\mr n F^2/9$ being proportionate to the fluctuations of
the  boundary. Expressions for the two terms of Eq.~\ref{eq:sigma2} must be
corrected for small forces since the properties of the desorbed tail are then
strongly affected by the presence of the wall. We do this in an \textit{ad hoc}
way to match the known behavior at $F=0$. Namely, the first term is amended by
adding an $F$-independent constant
\begin{equation}\label{eq:sigma1}
\sigma_1^2(Z)=\frac{\bar n}{3}+\bar n_0\Big(\frac{4-\pi}{6}-\frac{1}{3}\Big),
\end{equation}
where $\bar n_0=\bar n|_{F=0}$, so that at zero force the fluctuations of an
ideal mushroom are described correctly. It is clear that the second term,
$\sigma_2^2(Z)=\sigma^2_\mr n F^2/9$, vanishes at $F=0$. This underestimates
the effect of $n$ fluctuations since the mean end height in the mushroom state
never turns to 0. The correct  value of the second term at $F=0$ can be
inferred from the theory \cite{unusual1,unusual2,UPT_first_order}, and the
amended expression interpolates between the two limits of large and vanishing
force
\begin{equation}\label{eq:sigma22}
\sigma_2^2(Z)=\sigma^2_\mr n\left[\frac{F^2}{9}+\frac{\pi-2}{6\bar n_0}\exp\Big(-\frac{\bar nF^2}{6}\Big)\right].
\end{equation}
Here the interpolation is introduced \textit{via} an exponentially decaying
weight where the exponent, $\bar nF^2/6$, represents the stretching free energy
of the tail and quantifies the relative importance of the stretching effect in
comparison to that of the impenetrable wall.  The linearized expression for
$\delta$ is taken from the continuum ideal chain model:
{$\delta=\frac{F^*(F-F^*)}{3}$ \cite{UPT_first_order}

We go all this way in order to be able to utilize the effective interface model in describing the chain dynamics: accurate 
decomposition of total $Z$-fluctuations into two terms, one being related to the boundary fluctuations, and the other 
describing fluctuations of $Z$ at fixed boundary position, is central to the dynamic theory. \RE{The} theoretical prediction for 
chain end fluctuations is in good agreement with the simulation results (see Fig.(\ref{VZ})). This includes the power-law 
growth upon approaching the transition point, as well as the anomalous fluctuation $\sigma^2\propto\sigma^2_\mr n\propto N^2$ 
at the transition point $\delta=0$, consistent with the more sophisticated statistical 
theory \cite{unusual1,unusual2,UPT_first_order} and the data in Fig.(\ref{VZ}). The fact that the apparent exponent 
at the transition point $\sigma\propto N^{1.8}$ is slightly less than 2 reflects finite-size corrections to the dominant 
scaling in the explored range of $N$.

\section{Dynamic relation and critical slowdown}

We now apply the effective interface
model to estimate the relaxation dynamics in our system. We suppose that the
interface undergoes a random process, following a Langevin equation
\begin{equation}\label{eq:Langevin}
\frac{dn}{dt}=D(n)\cdot f(n) + \sqrt{2 D(n)} \: \eta, 
\end{equation}where $D(n)$ and $f(n)=-\frac{\partial\mc H_\mr{eff}}{\partial n}$ are the
mobility and the ``driving force'' and $\xi$ is a Gaussian distributed white
noise with variance $\langle \eta(t) \eta(t')\rangle = \delta(t-t')$
\cite{Gardiner}. The inverse mobility $D^{-1}(n)$ is governed by two contributions to the dissipation upon moving the boundary. 
First, there is viscous drag as the desorbed tail moves against the solvent, which scales as $\propto N$ for free-draining 
chains \cite{Doi_Edwards,UPT_first_order} and should be independent of the adsorption strength. Second, the motion of the 
boundary involves breaking the contact with the adsorbing surface. This source of dissipation is expected to depend 
on $\varepsilon$ but not on the tail length. Altogether, we write
\begin{equation}
\label{eq:D_n}
D^{-1}(n)=\alpha n\zeta_0+\zeta_\mr{ads},
\end{equation}where $\zeta_0$ is the translational monomer friction coefficient (in our dynamic simulation scheme $\zeta_0$ 
defines the time scale and is taken as 1), $\alpha$ is the fraction of the tail effectively involved in the boundary 
fluctuations, and $\zeta_\mr{ads}$ is the friction coefficient associated with local adsorption-desorption kinetics. 
A variable diffusion coefficient $D(n)$ leads to a multiplicative noise, which can be solved by a standard transformation of variables 
(see Supplementary material). This has a disadvantage of involving a cubic equation for the position of the relevant minimum.
To keep the analytics tractable we adopt a naive way of treating the diffusion coefficient as a constant evaluated at the 
most probable value of the tail length, $D=D(\bar n)$ where $\bar n$ is given by Eq. (\ref{eq:tail_length}).

We assume that the characteristic time scale for the motion of the boundary is
dominated by the diffusive process around the minimum of $\mc H_\mr{eff}(n)$ at
$\bar n$. Hence the evolution of $n$ is approximated by the Ornstein-Uhlenbeck
process, which has the characteristic relaxation time \cite{Risken}
\begin{equation}
\label{eq:taun}
\tau_n=D^{-1}(\bar n)\sigma^2_\mr n.
\end{equation}
Specifically, in the three limits $\tau_n \propto 1/|\df|^3$ for $\df < 0$ (adsorbed
state), $\tau_n \propto N/\df^2$ for $\df > 0$ (stretched state), and $\tau_n
\propto N^3$ for $\df =0$ (transition point). 

\begin{figure}[t]
\centering
\includegraphics[scale=0.45]{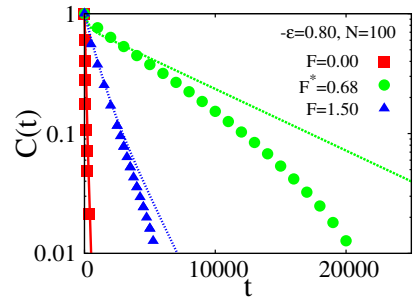}

\caption{\RE{Examples of time autocorrelation function $C(t)$ of the end 
position for chain length $N=100$ at $-\varepsilon=0.80$ and
three different stretching forces corresponding to an adsorbed state
($F < F^*$), a stretched state ($F > F^*$), and the transition ($F = F^*)$. 
Symbols correspond to data from BD simulations, and lines show the prediction 
from Eq.~(\ref{eq:C}) for $\zeta_0=1, \alpha=0.25,$, and $\zeta_{\mbox{\tiny ads}}=1.8$. } }
\label{fig:C}

\end{figure}

To calculate the autocorrelation function of the end position, $C(t) =
\frac{1}{\sigma^2(Z)} \big(\langle Z(t) \: Z(0) \rangle - \langle Z \rangle^2
\big)$, we rewrite $Z$ as $Z = \bar{Z}(n(t)) + y(t)$, where $y(t)$ accounts for
the conformational fluctuations of the desorbed tail at a fixed position of the
boundary. The autocorrelation function at fixed boundary is written as $\langle
y(t) y(0) \rangle \approx \sigma^2_1 \exp(-t/\tau_\mr R(\bar n))$ where
$\sigma_1^2$ is evaluated according to Eq.~(\ref{eq:sigma1}) and $\tau_\mr
R(\bar n)$ is the relaxation time of the end position at \RE{fixed} tail
length. This can be obtained as the average end-to-end distance relation time
of the multi-mode Rouse model with one end fixed: $\tau_\mr R(\bar
n)=\RE{\frac{1}{9}}\zeta_0\bar n^2$ \cite{Doi_Edwards}. Neglecting
cross-correlations we obtain 
\begin{equation} \label{eq:C}
C(t) = \frac{1}{\sigma_1^2(Z)+\sigma_2^2(Z)}\Big(\sigma_1^2(Z)e^{-t/\tau_\mr R(\bar n)} + \sigma_2^2(Z)e^{-t/\tau_\mr n} \Big).
\end{equation}
where the variances $\sigma_1^2(Z)$ and $\sigma_2^2(Z)$ are evaluated at the
most probable value $\bar n$ according to Eqs.(\ref{eq:sigma1},
\ref{eq:sigma22}). \RE{From $C(t)$ we can calculate the average relaxation time
\cite{simple_liquid},  $\tau = \int_0^\infty \!\! \ud t \: C(t)$, giving}
\begin{equation}
\label{tau_avr}
\tau = \frac{1}{1+\Phi} \tau_n + \frac{\Phi}{1+\Phi}\tau_\mr R(\bar n)
\end{equation}
where $\Phi=\sigma_1^2(Z)/\sigma_2^2(Z)$.  

\RE{Fig.~(\ref{fig:C}) compares simulation results for the autocorrelation
function with the theoretical predictions for $\alpha = 0.25$ and
$\zeta_\mr{ads}=1.8$ for selected parameter sets.  The data are in reasonable
agreement with the simulation results at early and intermediate times.
Deviations can be observed at late times especially at $F^*=0.68$, where the
simulation data decay faster than exponentially. In this regime, the boundary
diffuses along the whole chain, and the harmonic approximation (Eq.~\ref{eq:taun})
becomes questionable. However, although the differences in the figure for
curves at $F^*$ seem large, they appear mainly at late times where $C(t)$ is
small, and do not have a large effect on the average relaxation time $\tau$
(only about 20 \%). }

\RE{The simulation results for the average relaxation time, $\tau$ are shown in
\figref{dynamicslowdown} together with the theoretical predictions (again using
$\alpha=0.25$ and $\zeta_\mr{ads}=1.8$). In the fully desorbed state, $\tau$ is
dominated by $\tau_R(\bar{n})$ and proportional to $N^2$, and Rouse relaxation
describes the strong force limit very accurately. In the adsorbed state, $\tau$
is dominated by $\tau_\mr n$ and chain length independent, and it shows the
predicted power law dependence as one approaches the critical stretching point.
Hence we observe critical slowing down, unusual for a first order transition.
Right at the transition, the relaxation times become very slow and scales as
$\tau \propto N^{2.8}$ (\figref{dynamicslowdown}, inset), which is also close
to the limiting behavior predicted by the model, $\tau \sim \tau_\mr n \propto
N^3$.  Altogether, the dynamic model successfully captures three different
regimes (adsorbed, desorbed, and near-critical) for chains of different
lengths.}

\begin{figure}[h]
\centering
\includegraphics[scale=0.45]{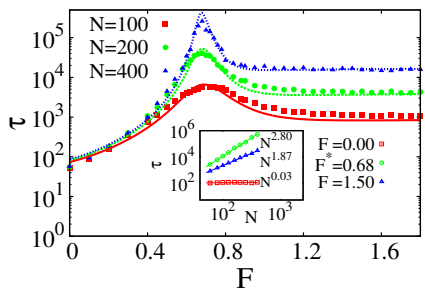}

\caption{Characteristic relaxation time $\tau$ of free end correlations
vs.\ stretching force for chain lengths
$N=100,200,400$ at $-\varepsilon=0.80$. Symbols are data from BD
simulations, and lines correspond to the prediction based on the dynamic model.
at $\zeta_0=1, \alpha=0.25,$, and $\zeta_{\mbox{\tiny ads}}=1.8$.  
Inset shows scaling of $\tau$ with $N$ in the adsorbed state
($F = 0$), at the transition ($F=F^*$), and in the stretched state
($F=1.5$). }
\label{dynamicslowdown}

\end{figure}

\section{Summary}

To summarize, we have utilized BD simulations to study the
adsorption-stretching transition of a polymer chain tethered onto an adhesive
solid surface, with a tension force exerted at the free chain end. 
Even though the transition is first order, we found critical (power-law) slowdown
close to the transition point, as is typical for continuous phase
transitions. Our observations could be rationalized by a simple
model that describes the system in terms of single interface
separating the adsorbed and stretched chain blocks. Close
to the transition, the relaxation dynamics is governed by the
time scale of the interface diffusion, which diverges at the
transition.

\RE{The good correspondence of the theory and simulations in the case of the
dynamic properties is perhaps fortuitous, since extra fitting parameters
related to friction coefficients were introduced in the theory.  However,
regarding the static properties, the choice of the amplitude parameter was
based on comparison between hard walls and logarithmic potential which was not
just a simple fitting of the final data.}

\RE{In the BD simulations, we adopt a coarse-grained off-lattice model, which
can be seen as the dynamic version of the off-lattice MC scheme proposed by
Laradji et al (see ref.\cite{CG_MC}). In this approach the excluded volume
interactions are replaced by soft interactions that are formulated in terms of
local monomer densities.  Compared to the more commonly used coarse-grained
models with non-bonded hard core potentials, this approach has the two
advantages.  First, it can be simulated very efficiently, since the most time
consuming part, an explicit evaluation of the pair potentials, is avoided.
Second, since it uses soft potentials, equilibrium times are comparatively
short. Soft potentials account properly for the density correlations (at least
on the large scales) that are characteristic of the excluded volume
interactions.  With such models, the static behavior of single chains (in
particular, the Flory exponent) can be captured correctly.  On the other hand,
they do not impose the dynamic constraint that chain self-crossing is
forbidden. Thus, topological effects such as entanglements or knots which may
influence the dynamics strongly, are not accounted for correctly here. For
single-chain dynamics in the presence of a stretching force these topological
effects are typically neglected.  This can be motivated by the observation that
single chains in good solvents typically do not contain knots at the chain
lengths considered in the present study \cite{virnau}. In the presence of a
stretching force, the knot probability should be further reduced.  Correlated
fluctuations due to excluded volume effects are important at zero (or very
weak) external forces.  Strong stretching generally screens out the global
excluded volume effects.  }

\RE{In the present work, we have demonstrated anomalous dynamic slowdown in the
vicinity of a first order transition. An additional source of slowdown comes
into play if one approaches the critical adsorption point (the point
$\epsilon^* = -0.42, F^*=0$ in Fig.\ 2). Whereas the slow dynamics in the
situation discussed here is a result of the slow diffusion of an interface,
critical slowdown is associated with the slow relaxation of very large loops.
These phenomena will be analyzed in more detail in future work.}

Our simple model might be useful also for the interpretation of other,
similar, macromolecular transitions, such as, e.g., the force-induced unzipping
transition of DNA or RNA \cite{DNA,DNA_model,DNA_finiteSize,unzipping_DNA}. In
simulations, the (nonequilibrium) transition times for unzipping were found to
scale as $N^3$ with the chain length \cite{DNA}, which is consistent with our
findings.  We would expect that the relaxation times in the vicinity of the
unzipping transition also show a critical power-law behavior. Unusual static
and dynamic behavior close to macromolecular phase transitions might be
ubiquitous in nature, therefore the study of such phenomena is more than just
intriguing and may bring new insight in understanding biological processes.

\tb{Supplementary material} Supporting Material contains the detailed description of the simulation scheme, a comparison of the two models (hard walls vs. logarithmic boundary potential) and the accurate treatment of the Langevin equation for the boundary with a variable diffusion coefficient.

\begin{center}
{\bf Acknowledgements}
\end{center}

This work has been supported by the German Science Foundation (DFG) within the
Graduate School of Excellence Materials Science in Mainz (MAINZ), the SFB TRR
146 (project C1) and the Grant Schm 985/13-2. S.Z.  acknowledges financial
support from the National Science Foundation of China (NSFC) 21374011,
21434001. Simulations were carried out on the computer cluster Mogon at JGU
Mainz (hpc.uni-mainz.de).

%


\begin{thebibliography}{36}
\bibitem [{not()}]{note1}%
  \BibitemOpen
  \href@noop {} {}\bibinfo {note} {{T}he analytical theory deals with the
  Gaussian chain model, where the excluded volume interactions are ignored. We
  found that the inclusion of excluded volume interaction mainly shifts the
  transition point.}\BibitemShut {Stop}%
\end{thebibliography}

\end{document}